\documentclass[10pt,twoside,letterpaper]{article}

\usepackage{amsmath}
\usepackage{mathtools}
\usepackage[margin=0.75in]{geometry}
\usepackage{graphicx}
\usepackage{authblk}
\usepackage{url}
\usepackage[hidelinks]{hyperref}
\author[1,2,3,*]{John L. Rubinstein}
\author[4]{Marcus A. Brubaker}
\affil[1]{Molecular Structure and Function Program, The Hospital for Sick Children}
\affil[2]{Department of Medical Biophysics, University of Toronto}
\affil[3]{Department of Biochemistry, University of Toronto}
\affil[4]{Department of Computer Science, University of Toronto}
\affil[*]{correspondence to: john.rubinstein@utoronto.ca}

\title{Alignment of cryo-EM movies of individual particles by optimization of image translations}
\begin{document}
\maketitle
\begin{abstract}
\noindent Direct detector device (DDD) cameras have revolutionized single particle electron cryomicroscopy (cryo-EM). In addition to an improved camera detective quantum efficiency, acquisition of DDD movies allows for correction of movement of the specimen, due both to instabilities in the microscope specimen stage and electron beam-induced movement. Unlike specimen stage drift, beam-induced movement is not always homogeneous within an image. Local correlation in the trajectories of nearby particles suggests that beam-induced motion is due to deformation of the ice layer. Algorithms have already been described that can correct movement for large regions of frames and for \(>1\) MDa protein particles. Another algorithm allows individual \(<1\) MDa protein particle trajectories to be estimated, but requires rolling averages to be calculated from frames and fits linear trajectories for particles. Here we describe an algorithm that allows for individual \(<1\) MDa particle images to be aligned without frame averaging or linear trajectories. The algorithm maximizes the overall correlation of the shifted frames with the sum of the shifted frames. The optimum in this single objective function is found efficiently by making use of analytically calculated derivatives of the function. To smooth estimates of particle trajectories, rapid changes in particle positions between frames are penalized in the objective function and weighted averaging of nearby trajectories ensures local correlation in trajectories. This individual particle motion correction, in combination with weighting of Fourier components to account for increasing radiation damage in later frames, can be used to improve 3-D maps from single particle cryo-EM.
\end{abstract}

\section{Introduction}
\noindent The use of CMOS technology in direct detector device (DDD) cameras for electron cryomicroscopy (cryo-EM) has enabled the acquisition of exposure series `movies'. Movies of radiation sensitive specimens revealed that beam-induced motion blurs images \cite{mcmullan2008electron,glaeser2011images,Brilot:2012zr}. DDD movies are typically acquired with exposures of 1 to 3 e\(^{-}\)/\AA\(^{2}\)/frame on the specimen, which corresponds to 2 to 5 e\(^{-}\)/pixel/frame on the detector, depending on microscope magnification. These low exposures result in low signal-to-noise ratios (SNRs) in individual movie frames. Optimal extraction of high-resolution information from images of single particles requires alignment of movie frames, a process that is complicated by the low SNR. Fig. 1A shows the average of a 30 frame movie acquired with 2.5 e\(^{-}\)/pixel/frame, corresponding to 1.2 e\(^{-}\)/\AA\(^{2}\)/frame on the specimen at 200 kV with a K2 Summit DDD (Gatan Inc). A few representative particles are circled in red. Fig. 1B shows a single frame from the movie, illustrating the low SNR of the frames. Frame alignment is complicated further by the presence of fixed pattern noise in images from errors in sensor gain normalization. Significant progress in image analysis has already been enabled by programs to perform rigid body translational alignment of entire field-of-view movie frames (currently 4000 \(\times \) 4000 pixels for most cameras). A method introduced by Li and colleagues \cite{Li:2013kq} decreases the weight of high spatial frequencies in images to suppress artifacts from fixed pattern noise before calculating pairwise cross-correlation functions between movie frames. The optimal frame displacement values from the cross-correlation functions are used to create a system of over-determined linear equations. Matrix algebra is then used to determined the frame-to-frame translations that best fit the data in a least squares sense. This least squares whole frame alignment method has allowed high-resolution structures to be determined for important biological macromolecules \cite{Li:2013kq,Cao:2013lq,Liao:2013db}.\\ 

\begin{figure}[ht]
\centering
  \includegraphics[width=0.75\textwidth]{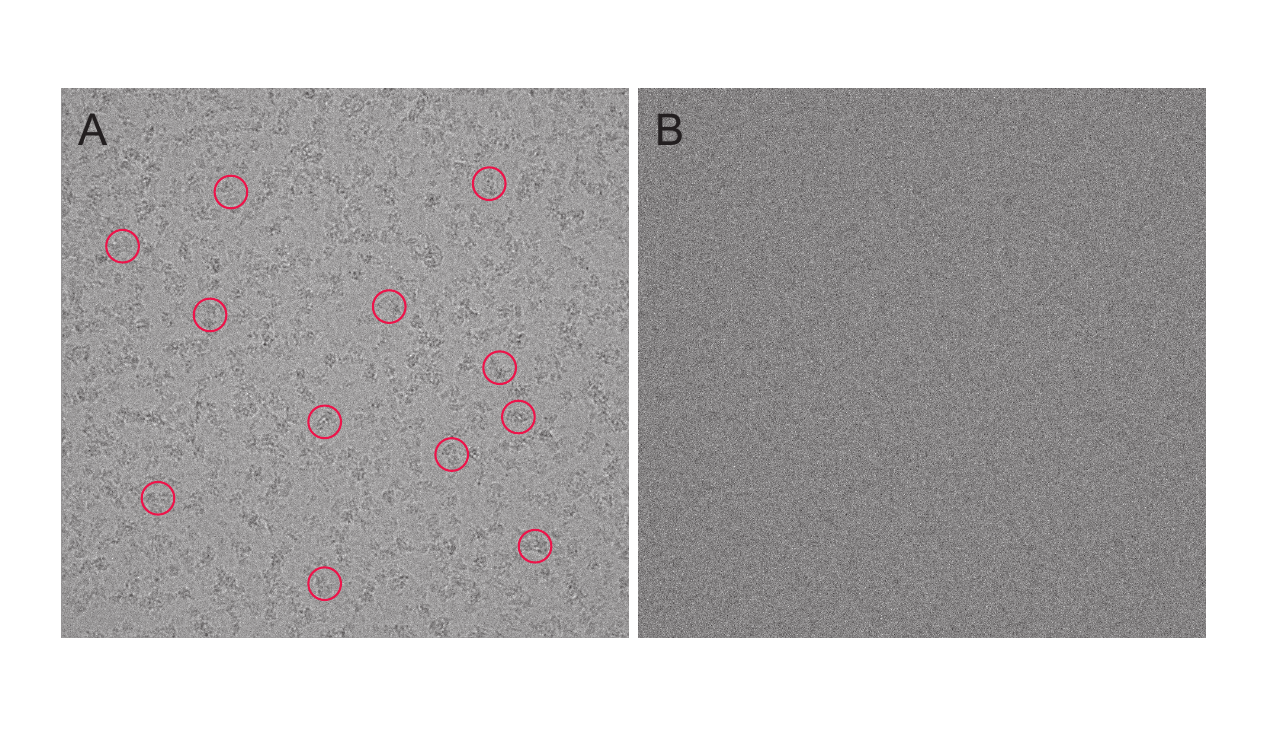}
   \caption{\textbf{Example micrograph. A,} The average of 30 frames after least squares alignment of whole frames for an exposure series movie of the \textit{Saccharomyces cerevisiae} V-ATPase in ice. The exposure used 2.5 \(e^{-}/pixel/frame\) and each pixel corresponds to 1.45 \(\times\) 1.45 \AA. The \textit{S. cerevisiae} V-ATPase has a molecular weight of 900 kDa. Several example complexes are circled in red. \textbf{B,} An individual frame from the movie shows the low SNR in frames.}
\end{figure}

\noindent Cryo-EM of large particles with DDDs has shown that beam-induced motion cannot be described completely by rigid body translation of entire movie frames \cite{Brilot:2012zr}. Instead, these experiments suggested that the beam-induced movement of ice-embedded protein is better described by a translation of each particle in each frame. Examination of tilt pairs of images demonstrated that rotation of specimens, probably due to movement of the ice layer, does occur \cite{Henderson:2011mz}. However, the magnitudes of these rotations were small and will have the most significant effect on particles with large radii, like viruses. The consequences of specimen rotation can be neglected at present without limiting map resolution for particles smaller than 1 MDa. The translation of particles in frames can be written as \(-\vec{t}_{z}=-(x_{z},y_{z})\) for each particle in frame \(z\), where \(-x_{z}\) and \(-y_{z}\) are the difference in particle position between frame \(z\) and frame \(1\) in the \(x\) and \(y\) directions, respectively. If these translations are known, their inverse (\(\vec{t}_{z}\)) can be applied to the particle images before averaging of frames to optimize the extraction of high-resolution information from the image. It is likely that accurate individual particle motion correction could extract information from images that is neglected by whole frame alignment. Despite the success of the least squares method for whole frame alignment, it was pointed out by the authors of the method that it is not reliably able to align image regions smaller than 2000 \(\times \) 2000 pixels for movies acquired using typical conditions. As such, the least squares method is not capable of aligning regions of frames that contain individual particles in order to correct for deformation of the ice layer during imaging. A method to align individual particle images was introduced that is tightly integrated into the single particle orientation estimation framework of the program \textit{Relion} and has resulted in several high-resolution structures \cite{Scheres:2012yu,Scheres:2014ys,campbell20152}. For small particles this approach requires rolling averages of frames, which increases the SNR over individual frames but loses information about true trajectories. Also, the individual particle trajectories for small particles from this method include errors, and it is necessary to fit linear trajectories with constant velocities for particles, which are not necessarily a good approximation for their true trajectories. Furthermore, the approach cannot readily be used outside of the \textit{Relion} software package.\\

\noindent Here we aim to identify the translations \(\vec{t}_{z}\) for movies of individual ice-embedded particles that best bring the frames into alignment for each particle, without the use of rolling frame averages or fitted linear trajectories. In order to produce a robust and computationally efficient method for correcting the effects of beam-induced movement in small regions in images, or on individual small (\(<1\) MDa) particles, we pose the problem in terms of optimization. We propose an objective function based on the correlation of the Fourier transforms of individual frames with the sum of all frames. A well-established iterative optimization algorithm that makes use of partial derivatives of the objective function is then used to find the desired translation values. Once optimized, this objective function gives frame-to-frame trajectories for images of individual particles that show strong local correlation. We show that smoothing of trajectories for individual particles can be used to identify and correct beam-induced particle movement. This approach, in combination with compensation for the fading of Fourier components due to radiation damage, was implemented in a new program that we call \textit{alignparts\_lmbfgs}.\\

\section{Methods and Results}

\subsection{Choice of objective function}
\noindent Based on the observation that averages of unaligned particle frames appear blurred, a reasonable alignment for each region of the frame that contains a particle is the alignment that makes the sum of all of the frames best agree with each of the frames. Accordingly, we propose an objective function that maximizes the sum of the correlations of the Fourier transform of each shifted frame with the sum of the Fourier transforms of the shifted frames. Prior to analysis, we apply a temperature factor in Fourier space with the form \(exp(\frac{-B}{4d^{2}})\) to prevent fixed pattern noise from dominating the analysis \cite{Li:2013kq}. The effect of translation on the Fourier transform of a movie frame is a phase change, \({\phi_{jz}}\), in each Fourier component of the frame, written \(F_{jz}\) for the \(j^{th}\) Fourier component of frame \(z\). The phase shifted Fourier component is given by \(F_{jz}(\cos{\phi_{jz}}+i\sin{\phi_{jz}})\) or \(F_{jz}S_{jz}\) where \(S_{jz}=(\cos{\phi_{jz}}+i\sin{\phi_{jz}})\). The amount of phase change is given by

\begin{equation}
\label{phase}
\phi_{jz}=k_{x}(j)\cdot  x_{z} \frac{2\pi}{N}+k_{y}(j)\cdot y_{z} \frac{2\pi}{N}
\end{equation}
\noindent where  \(N\) is the extent in pixels in both the \(x\) and \(y\) direction of the \(N \times N\) image, and \(k_x(j)\) and \(k_y(j)\) are the distance of the \(j^{th}\) Fourier component from the origin in the \(k_{x}\) and \(k_{y}\) directions, respectively. As described above, \(-x_{z}\) and \(-y_{z}\) are the difference in particle position between frame \(z\) and frame \(1\) in the \(x\) and \(y\) directions, respectively. The Fourier transform of a sum is equal to the sum of Fourier transforms. Consequently, the \(j^{th}\) Fourier component from the sum of the shifted frames of a movie with \(Z\) frames is given by \(\sum_{z=1}^{Z}F_{jz}S_{jz}\). The unnormalized correlation between two Fourier transforms, \(F_{1}\) and \(F_{2}\), is given by \(F_{1} \cdot F^{*}_{2}\) where \(*\) denotes the complex conjugate. For the correlation between the sum image and the individual frame, these values must be summed for the \(J\) Fourier components in a resolution band \(\vec{k}(j) \in [\vec{r}_{min},\vec{r}_{max}]\). It is only necessary to consider two times the real part of the expression for the correlation, because the Fourier transforms of real functions, such as images, are Hermitian, so that for every term in the correlation \((a_{1}+b_{1}i)(a_{2}-b_{2}i)=a_{1}a_{2}+b_{1}b_{2}+(a_{2}b_{1}-a_{1}b_{2})i\) there is a corresponding term \((a_{1}-b_{1}i)(a_{2}+b_{2}i)=a_{1}a_{2}+b_{1}b_{2}-(a_{2}b_{1}-a_{1}b_{2})i\) and adding these two terms removes their imaginary parts. In an objective function, the factor of 2 may be neglected without changing the position of the optimum, and the negative of the function can be used in order to interface with pre-existing optimization algorithms, which typically seek to minimize functions. Consequently, we can propose an objective function, \(O(\Theta)\), that meets the criterion described above:

\begin{equation}
\label{fom}
\begin{aligned}
O(\Theta)=-Re\sum_{z=1}^{Z} \sum_{j=1}^{J}\left[ F_{jz}^{\mbox{*} }S_{jz}^ {\mbox{*} } \sum_{z^{\prime}=1}^{Z}F_{jz^{\prime}} S_{jz^{\prime}} \right]
\end{aligned}
\end{equation}

\noindent where \(\Theta\) represents the set of all \(x_{z}\) and \(y_{z}\) values, and \(S_{jz}^{\mbox{*}}=(\cos{\phi_{jz}}-i\sin{\phi_{jz}})\) is the complex conjugate of \(S_{jz}\), with \(\phi_{jz}\) calculated from \(x_{z}\) and \(y_{z}\) according to equation \ref{phase}. With equation \ref{fom} as the objective function, iterative optimization methods can be used to explore the \(\left(2\times Z\right)\)-dimensional space of frame translations to find values of \(x_{z}\) and \(y_{z}\) that minimize the function.

\subsection{Partial derivatives of the objective function}
\noindent Numerous algorithms exist for optimizing objective functions. Optimization problems can benefit greatly from the ability to analytically determine partial derivatives, or gradients, of the objective function with respect to all variables. The derivatives of \(S_{jz}\) and \(S_{jz}^{\mbox{*}}\) with respect to \(x_{a}\), the shift in the \(x\) direction for the \(a^{th}\) frame, are

\begin{equation*}
\frac {\partial S_{jz}}{\partial{x_{a}}}=(-\sin{\phi_{jz}}+i\cos{\phi_{jz}}) \frac{\partial \phi_{jz}}{\partial x_{a}}=iS_{jz}\frac{\partial \phi_{jz}}{\partial x_{a}}
\end{equation*}

\noindent and 

\begin{equation*}
\begin{aligned}
\frac {\partial S_{jz}^{\mbox{*}}}{\partial{x_{a}}}&=(-\sin{\phi_{jz}}-i\cos{\phi_{jz}}) \frac{\partial \phi_{jz}}{\partial x_{a}}=-iS_{jz}^{\mbox{*}}\frac{\partial \phi_{jz}}{\partial x_{a}}.
\end{aligned}
\end{equation*}

\noindent Using these simplifications, the derivative of the objective function in equation \ref{fom} with respect to \( x_{a}\) is

\begin{equation*}
\begin{aligned}
\frac{\partial O(\Theta)}{\partial x_{a}} 
&=-Re\sum_{j=1}^{J} \sum_{z=1}^{Z}
\left[ F_{jz}^{\mbox{*}} S_{jz}^ {\mbox{*}} \sum_{z^{\prime}=1}^{Z} F_{jz^{\prime}} \frac {\partial S_{jz^{\prime}}}{\partial x_{a}} +
F_{jz}^{\mbox{*}}\frac{ \partial S_{jz}^ {\mbox{*}}}{\partial x_{a} } \sum_{z^{\prime}=1}^{Z}F_{jz^{\prime}} S_{jz^{\prime}}  \right]\\
&=-Re\sum_{j=1}^{J} \left[ \sum_{z=1}^{Z}
\left( F_{jz}^{\mbox{*}}S_{jz}^ {\mbox{*}} \sum_{z^{\prime}=1}^{Z}F_{jz^{\prime}} iS_{jz^{\prime}}\frac{\partial \phi_{jz^{\prime}}}{\partial x_{a}} \right)
- \sum_{z=1}^{Z} \left(i F_{jz}^{\mbox{*}} S_{jz}^{\mbox{*}}\frac{\partial \phi_{jz}}{\partial x_{a}} \sum_{z^{\prime}=1}^{Z}F_{jz^{\prime}} S_{jz^{\prime}} \right) \right].\\
\end{aligned}
\end{equation*}

\noindent Noting that \(\partial \phi_{jz}/\partial x_{a} =0\) when \(a \neq z \) and \(\partial \phi_{jz}/\partial x_{a} =2 \pi k_{x}(j)/N\) when  \(a=z\) , the expression simplifies further:

\begin{equation}
\begin{aligned}
\label{fomderivativex}
\frac{\partial O(\Theta)}{\partial x_{a}}
&=-Re\sum_{j=1}^{J} \left[ \sum_{z=1}^{Z}
\left( iF_{jz}^{\mbox{*}}S_{jz}^ {\mbox{*}} F_{ja} S_{ja}\frac{2 \pi k_{x}(j)}{N} \right)
- \left(i F_{ja}^{\mbox{*}} S_{ja}^{\mbox{*}}\frac{2 \pi k_{x}(j)}{N} \sum_{z=1}^{Z}F_{jz} S_{jz} \right) \right]\\
&=-Re\sum_{j=1}^{J} \frac{2 \pi ik_{x}(j)}{N} 
\left[ F_{ja} S_{ja} \sum_{z=1}^{Z} F_{jz}^{\mbox{*}}S_{jz}^ {\mbox{*}}
- F_{ja}^{\mbox{*}} S_{ja}^{\mbox{*}}\sum_{z=1}^{Z}F_{jz} S_{jz}  \right].\\
\end{aligned}
\end{equation}

\noindent Similarly, the partial derivative of equation \ref{fom} with respect to  \(y_{a}\) is

\begin{equation}
\label{fomderivativey}
\frac{\partial O(\Theta)}{\partial y_{a}} =-Re\sum_{j=1}^{J} \frac{2 \pi ik_{y}(j)}{N} 
\left[ F_{ja} S_{ja} \sum_{z=1}^{Z} F_{jz}^{\mbox{*}}S_{jz}^ {\mbox{*}}
- F_{ja}^{\mbox{*}} S_{ja}^{\mbox{*}}\sum_{z=1}^{Z}F_{jz} S_{jz}  \right].
\end{equation}

\noindent We elected to use the limited memory Broyden-Fletcher-Goldfarb-Shanno (lm-bfgs) algorithm \cite{byrd1995limited} to optimize the objective function in equation \ref{fom}. By providing equations \ref{fom}, \ref{fomderivativex}, and \ref{fomderivativey} for lm-bfgs optimization, values of \(x_{z}\) and \(y_{z}\) were obtained for movies of V-ATPase particles in ice. Fig. 2A shows the calculated trajectories from optimization of 200 regions of 320 \(\times \) 320 pixels in each frame. These 200 image regions were selected by template matching from the image in Fig. 1A, and contain a mixture of usable particle images and other image features. The trajectories show local correlation, even though at this stage in the analysis individual particle trajectories are not provided with any information about the trajectories of nearby particles, except for any overlap in the 320 \(\times \) 320 pixel boxes. Close inspection of the trajectories in two regions of the micrograph (Fig. 2Bi and ii) reveals noise in the trajectories of individual particles obtained by the optimization method.

\begin{figure}[ht]
\centering
  \includegraphics[width=0.75\textwidth]{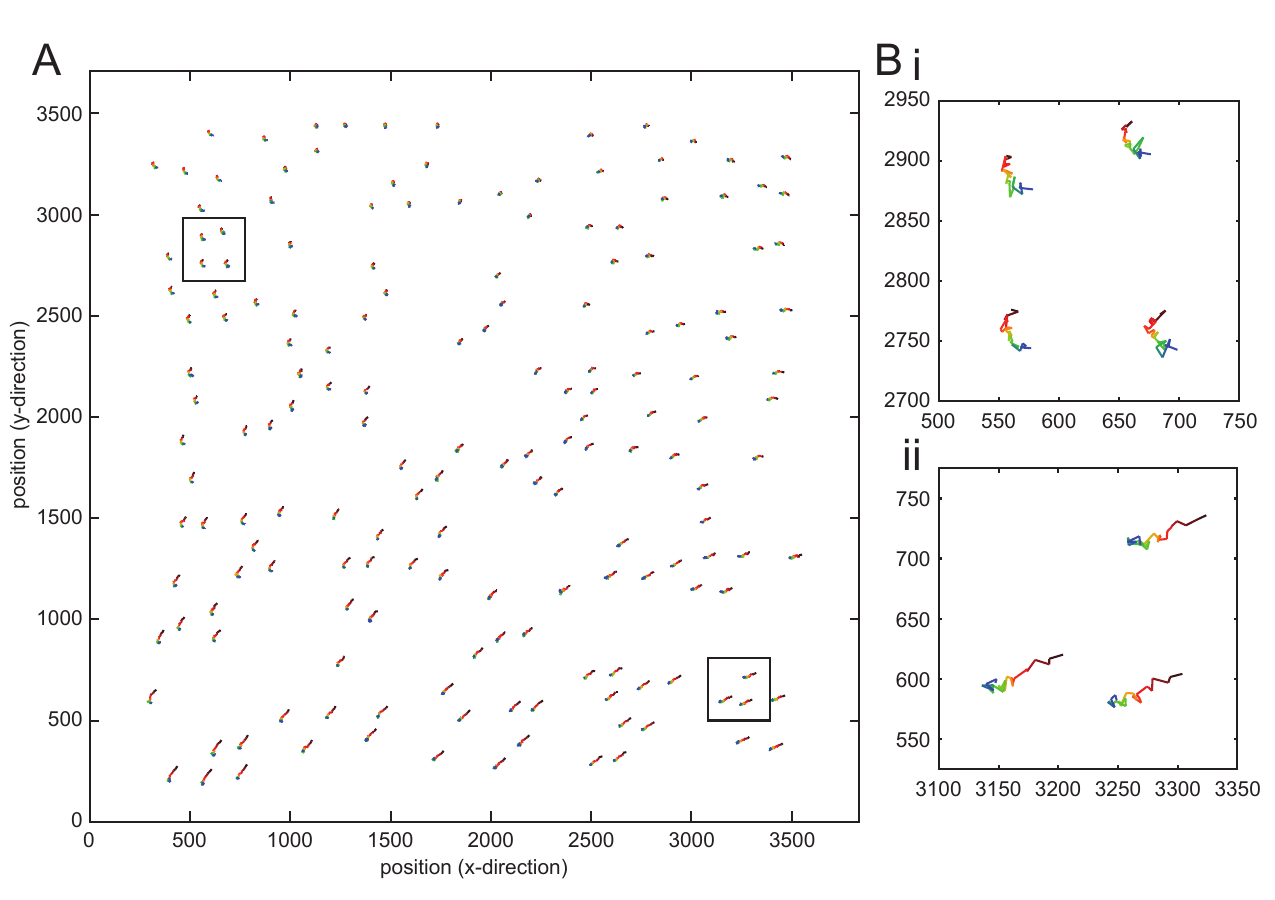}
   \caption{\textbf{Raw trajectories from lm-bfgs optimization of the objective function. A,} A plot of individual particle trajectories in image regions that are 320 \(\times\) 320 pixels. Each line in the plot indicates the trajectory of a single particle from frame 1 (black) to frame 30 (blue), exaggerated by a factor of 5. As can be seen from the plot, there is local correlation of particle trajectories. \textbf{B,} Inspection of individual particle trajectories from two regions of the micrograph (\textbf{i} and \textbf{ii}) reveals that there is significant noise in the trajectories.}
\end{figure}

\subsection{Smoothing}

\noindent Although encouraging, the noise seen in trajectories of particles in Fig. 2Bi and ii suggests that the optimization does not show the true trajectories of individual particle images. One obvious approach to reducing noise in a trajectory is to calculate the trajectory from a larger portion of the image, thereby increasing the signal available for calculating the objective function. Unfortunately, as the size of the box used for determining particle positions increases, particles must progressively be excluded that fall too close to the edge of the image. Increasing box sizes also results in almost identical trajectories for nearby particles that may mask the local variation in movement that this technique aims to recover. Better noise removal can be achieved by using two reasonable assumptions that are neglected in the analysis presented in Fig. 2. The first assumption is that trajectories are unlikely to have sudden changes in direction, although the possibility of these changes cannot be eliminated. The second assumption is that nearby particle trajectories are correlated. Enforcing these two conditions can be used to `smooth' particle trajectories to remove noise.

\begin{figure}[ht]
\centering
  \includegraphics[width=0.75\textwidth]{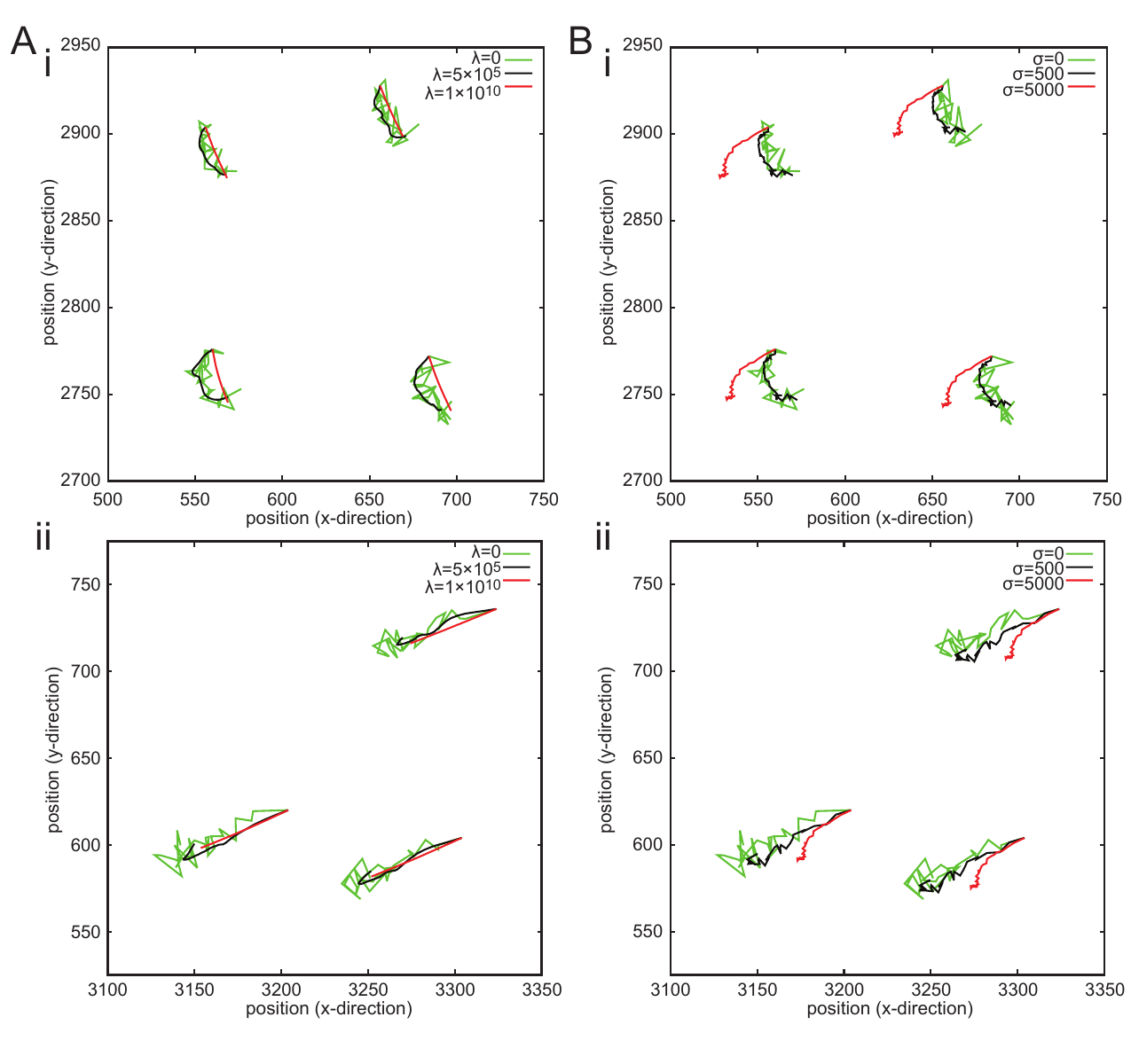}
  \caption{\textbf{Effects of two approaches for smoothing particle trajectories. Ai} and \textbf{ii,} Plots of trajectories in two small regions of the same micrograph in Fig. 2 show unsmoothed trajectories (green line), second-order smoothed with trajectories with sufficient smoothing \(\lambda=5 \times 10^{5}\) (black line), and second-order smoothed trajectories with excessive smoothing \(\lambda=1 \times 10^{10}\) (red). With excessive smoothing, trajectories have been forced to have a single linear drift rate. \textbf{Bi} and \textbf{ii,} Plots of trajectories in two small regions of the same micrograph showing unsmoothed trajectories (green line), locally averaged trajectories with a reasonable weighting \(\sigma=500\) (black line), and locally averaged trajectories with excessive weighting \(\sigma=5000\) (red). With excessive local averaging, trajectories in different parts of the micrograph have been forced to be identical. Particle trajectories are exaggerated by a factor of 5.}
\end{figure}

\subsubsection{Second order smoothing}

\noindent The assumption that true particle trajectories are unlikely to undergo sudden and dramatic changes in direction can be enforced by penalizing changes in \(\partial x_{z}/\partial z \) and \(\partial y_{z}/\partial z \). If \(\partial x_{z}/\partial z \) and \(\partial y_{z}/\partial z \) are constant (\(\partial^{2} x_{z}/\partial z^{2} \) and \(\partial^{2}y_{z}/\partial z^{2} \) are 0), the expected value for \((\vec{t}_{z}-\vec{t}_{z-1})\) is \((\vec{t}_{z-1}-\vec{t}_{z-2})\). Deviation from this expected linear trajectory can be penalized by an amount \(\lambda \left(\left[\vec{t}_{z}-\vec{t}_{z-1} \right]  -\left[\vec{t}_{z-1}-\vec{t}_{z-2}\right]\right)^{2}\). The overall penalty imposed on the objective function to encourage smoothness is then given by

\begin{equation}
\label{secondordersmoothing}
P(\Theta)=\sum_{z=3}^{Z} \lambda \left[ \left({x_{z}}-2{x_{z-1}}+{x_{z-2}}\right)^{2}+\left({y_{z}}-2{y_{z-1}}+{y_{z-2}}\right)^{2} \right]
\end{equation}
\noindent where \(\lambda\) is a user selected weighting parameter. This penalty is known as second order smoothing because it penalizes finite difference approximations of the second derivatives of \(x_{z}\) and \(y_{z}\) with respect to \(z\), \(\partial^{2} x_{z}/\partial z^{2} \) and \(\partial^{2}y_{z}/\partial z^{2} \). The penalty function described in equation \ref{secondordersmoothing} is added to the objective function in equation \ref{fom} to obtain the overall objective function that is optimized. The contribution to the penalty function in equation \ref{secondordersmoothing} from shifting of the \(a^{th}\) frame when \(a \in \left[3,Z-2\right] \) is\\
\(\lambda [(\vec{t}_{a}-2\vec{t}_{a-1}+\vec{t}_{a-2})^{2}+(\vec{t}_{a+1}-2\vec{t}_{a}+\vec{t}_{a-1})^{2}+(\vec{t}_{a+2}-2\vec{t}_{a+1}+\vec{t}_{a})^{2}]\) and consequently the first derivative of equation \ref{secondordersmoothing} with respect to \( x_{a}\) is given by

\begin{equation}
\label{smoothingderivativex}
\frac{\partial P(\Theta)}{\partial x_{a}} =
\begin{cases}
2\lambda  \left( x_{a} -2x_{a+1} + x_{a+2}  \right), & a=1,\\
2\lambda  \left( -2x_{a-1} +5x_{a} -4x_{a+1}+x_{a+2} \right), & a=2,\\
2\lambda  \left( x_{a-2}-4x_{a-1}+6x_{a}-4x_{a+1}+x_{a+2} \right), & a \in \left[ 3,Z-2 \right],\\
2\lambda  \left( x_{a-2} -4 x_{a-1}+5x_{a}-2x_{a+1}\right), & a=Z-1,\\
2\lambda  \left( x_{a-2} -2 x_{a-1} +x_{a} \right), & a=Z.
\end{cases}
\end{equation}

And similarly

\begin{equation}
\label{smoothingderivativey}
\frac{\partial P(\Theta)}{\partial y_{a}} =
\begin{cases}
2\lambda \left(y_{a} -2y_{a+1}+y_{a+2}  \right), & a=1,\\
2\lambda \left(-2y_{a-1} +5y_{a} -4y_{a+1}+y_{a+2} \right), & a=2,\\
2\lambda \left(y_{a-2}-4y_{a-1}+6y_{a}-4y_{a+1}+y_{a+2} \right), & a \in \left[ 3,Z-2 \right],\\
2\lambda \left(y_{a-2} -4y_{a-1}+5y_{a}-2y_{a+1}\right), & a=Z-1,\\
2\lambda \left(y_{a-2} -2y_{a-1}+y_{a} \right), & a=Z.
\end{cases}
\end{equation}

\noindent The derivative of the smoothed objective function is therefore the sum of the values from equations \ref{fomderivativex} and \ref{smoothingderivativex} for the derivative with respect to \(x_{a}\), and the sum of the values from equations \ref{fomderivativey} and \ref{smoothingderivativey} for the derivative with respect to \(y_{a}\). Fig. 3A shows the effect of increasing values of the user set parameter \(\lambda\) for two regions on opposite sides of the micrograph (Fig. 3Ai and ii). With \(\lambda=0\), the trajectories are noisy, as seen in Fig. 2. With \(\lambda=1 \times 10^{5}\) a significant amount of noise has been removed from the trajectories. Note also that nearby trajectories appear to be correlated even though this condition has not been enforced. With  \(\lambda=1 \times 10^{10}\), an excessively large number for these images, trajectories have been forced to become linear. Forcing trajectories to be linear is equivalent to fitting a single drift rate for each particle in the movie. The effects of different values of \(\lambda\) depend on the number of frames in a movie and the pixel values in those frames.

\subsubsection{Local averaging for smoothing}
\noindent Local correlation of nearby particle trajectories without the use of an increased box size can be achieved by weighted averaging after trajectories are calculated. In this approach, `raw trajectories' are determined for individual particle images with or without second order smoothing. Once raw trajectories are determined, locally averaged trajectories are calculated according to 

\begin{equation}
\label{neigboursmoothing}
\vec{t_{nz}}^{ \prime}=\frac{\sum_{m=1}^{M}w_{nm} \vec{t}_{mz}}{\sum_{m=1}^{M}w_{nm}}
\end{equation}
where \(\vec{t_{nz}}^{ \prime} \) is the smoothed displacement vector for the \(n^{th}\) particle in the \(z^{th}\) frame and \(\vec{t}_{mz}\) is the original displacement vector for the \(m^{th}\) particle in the \(z^{th}\) frame. The weight \(w_{mn}\) is given by 

\begin{equation}
\label{neigboursmoothingweight}
w_{mn}=exp \left(\frac{-d_{mn}^2}{2\sigma^2} \right)
\end{equation}
\noindent where \(d_{mn}\) is the distance between the \(m^{th}\) and \(n^{th}\) particles and \(\sigma\) is a user set parameter that determines the extent to which the smoothing is applied. This Gaussian weighting is equivalent to the local averaging used for fitting linear trajectories in \textit{Relion} \cite{Scheres:2014ys}. Because of the Gaussian form of equation \ref{neigboursmoothingweight}, 95 \% of of the weight for a particle trajectory will come from the trajectories within \(2\sigma \) pixels of that particle. Fig. 3B shows the effect of increasing the \(\sigma\) parameter for two sets of nearby trajectories (Fig. 3Bi and ii), without the use of second order smoothing. With \(\sigma=0\), the trajectories are noisy, as seen in Fig. 2. With \(\sigma=500\) a significant amount of noise has been removed from the trajectories, even though smoothness has not be enforced. With  \(\sigma=5000\), an excessively large number, trajectories on opposite sides of the micrograph from each other have been forced to be similar. In this situation, depending on the number of particles selected in the micrograph, the method becomes a nearly rigid frame alignment. Ideal smoothing of particle trajectories comes from combining the two approaches described above. Fig. 4A shows trajectories with \(\lambda=1 \times 10^{4}\) and \(\sigma=500\). As can be seen in two enlarged regions from opposite sides of the micrograph (Fig. 4Bi and ii), individual particle trajectories appear smooth with strong local correlation but significant variation from one edge of the micrograph to the other.

\begin{figure}[ht]
\centering
  \includegraphics[width=0.75\textwidth]{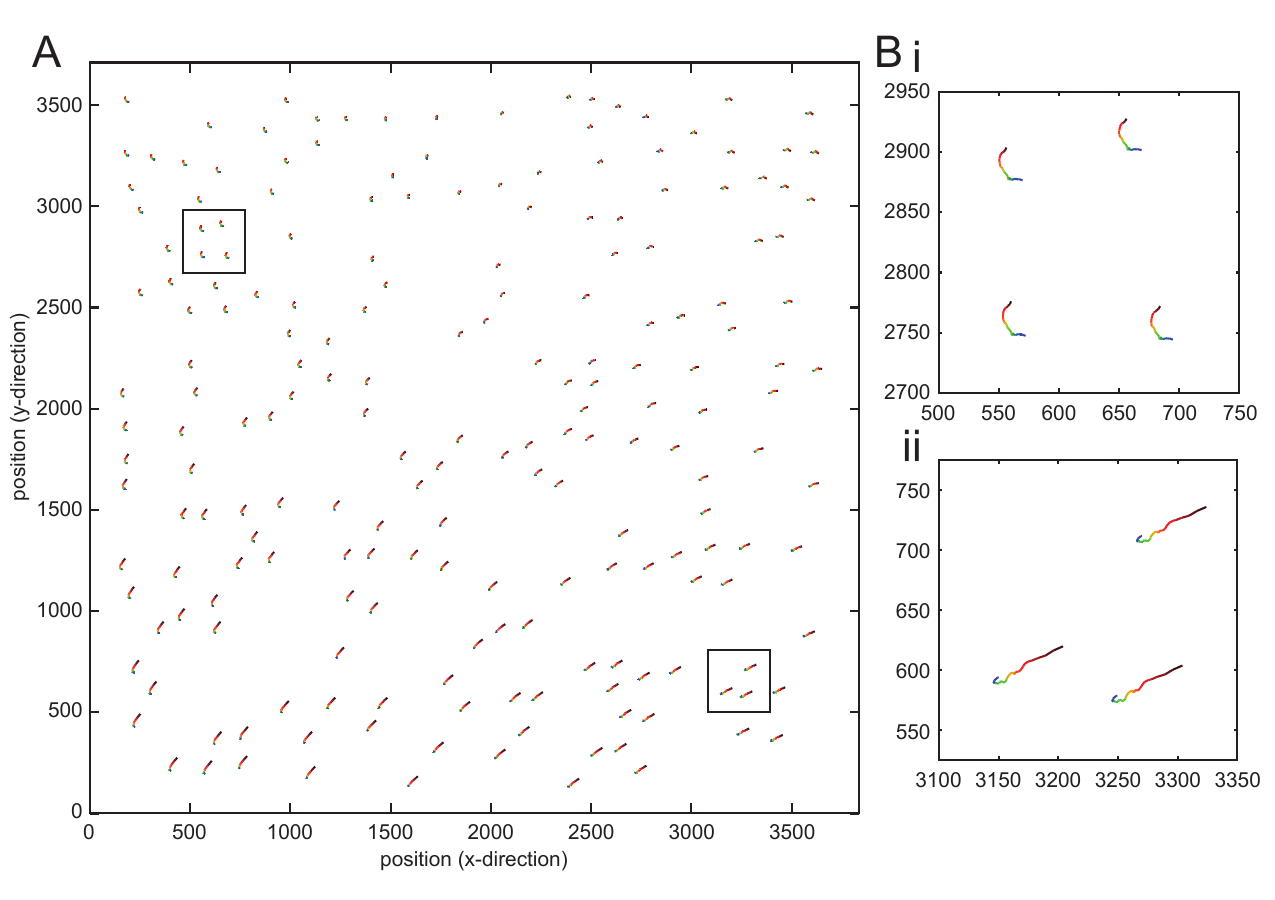}
  \caption{ \textbf{Combined smoothing approaches. A,} Particle trajectories for the whole micrograph and for two regions on opposite sides of the image (\textbf{Bi} and \textbf{ii}). By applying a combination of local averaging (\(\sigma=500\)) and second-order smoothing (\(\lambda=1 \times 10^{4}\)) individual particle trajectories display local uniformity but global variation and avoid sudden direction changes for trajectories. Particle trajectories are exaggerated by a factor of 5.}
\end{figure}

\subsubsection{Exposure weighting}
\noindent Correcting for the movement between frames restores the phases of each Fourier component to their correct values. 2-D crystal studies have shown that diffraction spots fade according to the relationship \(|\mathcal{F}_{j}(N)|^{2}=|\mathcal{F}_{j}(0)|^{2}e^{\frac{-N}{N_{e}(j)}}\), where \(|\mathcal{F}_{j}(0)|^{2}\) is the instantaneous intensity of the diffraction spot before electron exposure, \(|\mathcal{F}_{j}(N)|^{2}\) is the instantaneous intensity at exposure \(N\), and \(N_{e}(j)\) is the critical exposure at which the intensity of the diffraction spot fades to \(1/e\) times its initial intensity \cite{unwin1975molecular,hayward1979radiation}. The use of estimates of \(N_{e}(j)\) to optimize the combination of DDD movie frames was proposed previously \cite{Baker:2010ty}, with the necessary values measured from the fading of calculated diffraction spots from crystals. More recently these values were recorded for the same purpose from single particle analysis experiments \cite{grant2015measuring}. We included this exposure-weighting approach in \textit{alignparts\_lmbfgs}, utilizing \(N_{e}(j)\) estimates from the single particle data  \cite{grant2015measuring}. 

\section{Characterization of the algorithm}
To test the \textit{alignparts\_lmbfgs} algorithm, we compared the resolutions of maps calculated from a small dataset of images of 20S proteasome particles obtained with the same camera and microscope conditions described above for the V-ATPase (Latham, Benlekbir, and Rubinstein, in preparation). The resolution band used to align particle images was set at 500 to 40 \AA \ and the temperature factor to 2000 \AA \(^2\). The alignment of particle images was not sensitive to most changes in the low-resolution frequency cutoff, while including higher-resolution information in the alignment by changing the high-resolution frequency cutoff and decreasing the temperature factor could introduce noise into the particle trajectories. Particle images were subjected to 2-D classification, 3-D classification, and 3-D map refinement in \textit{Relion}. We compared whole frame alignment, individual particle motion correction with \textit{alignparts\_lmbfgs}, and individual particle motion correction including weighted averaging of Fourier components based on radiation damage. The resolutions that could be obtained (Fig. 5) were 4.17 \AA \ for whole frame alignment, 3.82 \AA \ for individual particle motion correction, and 3.68 \AA \ for individual particle motion correction with compensation for radiation damage. It is important to note that when using \textit{Relion}, the best resolution for a dataset is often obtained when the dataset is taken through the entire map calculation process, including 2-D classification and 3-D classification with each 3-D map here being constructed from approximately 14,500 particle images. Calculating a map from local motion corrected and exposure weighted images but selecting those images based on 2-D and 3-D classification of images from whole frame motion correction produced a map at 3.64 \AA. However, calculating a map from whole frame motion corrected images that had been selected based on 2-D and 3-D classification of local motion corrected and exposure weighted images gave a resolution of 4.32 \AA. The time required for the program to process images depends on the size of particle images, the number of frames in a movie, and the tolerance setting of the termination test within the lm-bfgs optimizer. Increasing second order smoothing by increasing the \(\lambda\) parameter also causes the algorithm to take more time to find the minimum in the objective function. However, a typical example consisting of particle images that are 256 \(\times\) 256 pixels with 30 frames per movie required approximately 0.3 s per particle image when running as a single process on a single core of an Intel i7-4790K CPU with a 4.0 GHz clock rate.\\

\begin{figure}[ht]
\centering
  \includegraphics[width=0.75\textwidth]{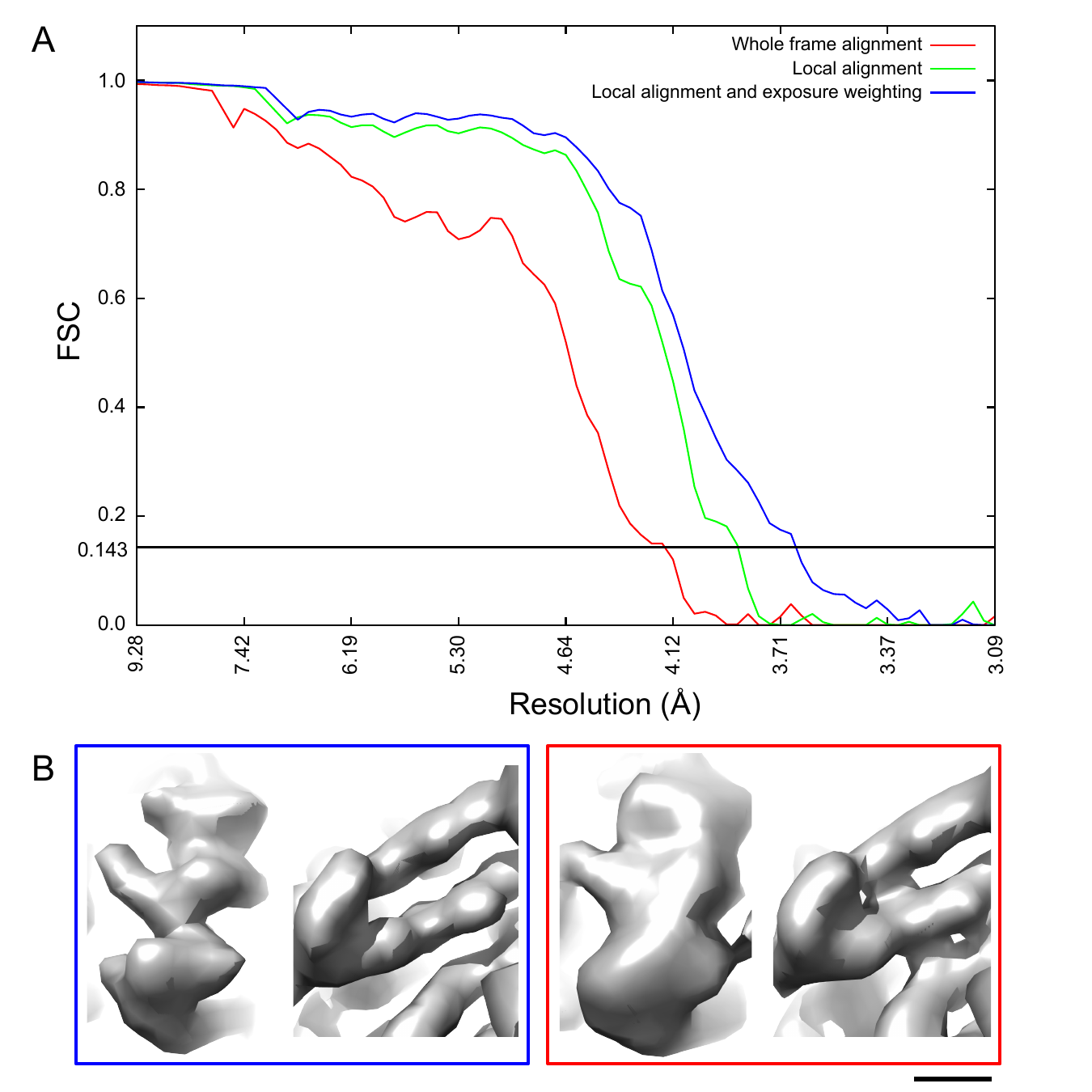}
  \caption{ \textbf{Comparison of frame alignment methods. A,} Fourier Shell Correlation curves show the resolution obtained for a test dataset using whole frame alignment only (red), alignment of particle images using the \textit{alignparts\_lmbfgs} individual particle motion correction algorithm (green), and using the individual particle motion correction algorithm including weighted averaging of Fourier components for radiation damage (blue). \textbf{B,} Regions from the individual particle motion corrected and exposure weighed map (blue box) showing an \(\alpha\)-helix (left) and \(\beta\)-turn (right) and the same regions from the map created using whole frame alignment only (red box). Scale bar, 5 \AA.}\end{figure}

\section{Discussion}

\noindent For full frame alignment, the least squares algorithm proposed by Li and colleagues possesses an advantage over the approach described here, in that the frame translations are highly over-determined: a movie consisting of \(Z\) frames will provide \((Z-1)\frac{Z}{2}\) equations that can be used to determine the \(Z-1\) frame translations needed to correct motion \cite{Li:2013kq}. However, while the least squares method correlates low SNR frames with other low SNR frames, \textit{alignparts\_lmbfgs}, or a whole frame implementation of the algorithm \textit{alignframes\_lmbfgs}, correlates low SNR frames with the relatively high SNR sum of frames. Consequently, the \textit{alignparts\_lmbfgs} method is able to work with image boxes at least as small as the 256 \(\times \) 256 pixel boxes used here and \textit{alignframes\_lmbfgs} is likely able to align lower-contrast whole frames than the least squares methods. The \textit{alignparts\_lmbfgs} approach should behave similarly to a simplistic iterative approach where frames are averaged and individual frames are subsequently aligned to the average. Special care must be taken in this simple iterative approach to ensure that the \(a^{th}\) frame is not aligned to an average where the \(a^{th}\) frame has been included at a fixed position, which could bias the alignment of the \(a^{th}\) frame. With \textit{alignparts\_lmbfgs}, the average always includes the \(a^{th}\) frame with the translations for the \(a^{th}\) frame that are being tested. Also, in \textit{alignparts\_lmbfgs} changing the translations for the \(a^{th}\) frame instantaneously affects the correlation of all other frames with the sum image, while with the simple iterative approach it does not, possibly making the identification of a global optimum less robust. The simple iterative approach will also almost certainly be slower than \textit{alignparts\_lmbfgs} at finding the optimum alignment of frames. The \textit{alignparts\_lmbfgs} approach benefits from being able to incorporate the second order smoothness constraint directly into the objective function. Both algorithms could become trapped in a local alignment minimum. However, the form of equation \ref{fom} suggests that the problem may be convex and in practice local minima do not appear to cause problems. The \textit{Relion} procedure \cite{Scheres:2014ys} integrates estimation of particle trajectories with projection matching from a reference map of the protein complex. Both procedures attempt to regularize particle trajectories: \textit{Relion} by using a running average of particle frames and fitting of a linear trajectory, \textit{alignparts\_lmbfgs} by introducing the second order smoothness constraint. The \textit{Relion} approach has the potential advantage that projections from a refined 3D map will posses stronger signal than the sum of all frames used as a reference in \textit{alignparts\_lmbfgs}. The potential disadvantage of the \textit{Relion} approach relative to \textit{alignparts\_lmbfgs} is that errors in contrast transfer function (CTF) estimation, structured noise in images from sample contamination or ice contamination, differing conformation of the protein particle in the image and map, and any other sources of inaccuracy in projection matching could affect the accuracy of trajectory estimation. Compared to the procedure introduced in \textit{Relion}, \textit{alignparts\_lmbfgs} is much less computationally expensive.\\

\noindent The two different smoothing approaches, second order smoothing and local weighted averaging of trajectories, have different advantages and uses. Second order smoothing is independent of particle density in images. If images contain few particles, their trajectories should be smoothed by increasing the second order smoothing parameter \(\lambda\). Local averaging of trajectories will have little effect for particles that are far apart from each other but can be applied effectively where there are many particles or other image features that can be aligned. For this situation, the amount of second order smoothing can be decreased somewhat. The value of the \(\lambda\) parameter that is used for aligning particle images depends on the value of the objective function in equation \ref{fom}. Therefore, for the same amount of smoothing \(\lambda\) will need to be bigger for images that have larger pixel values or more frames than for images with smaller pixel values or fewer frames. Consequently, \(\lambda\) should be tuned to produce a physically reasonable trajectory for particles by testing a variety of values with a small subset of the data. In practice, we have found that for consistent imaging conditions a fixed value of \(\lambda\) produces consistent results. In comparison, the \(\sigma\) value should be set to reflect how quickly trajectories are expected to vary across the image, which will depend on the magnification at which image were obtained. At higher resolution, or for larger particles, it could be useful to interpret trajectories to include the rotations of particles in the ice layer.\\

\section{Software and data availability}
The programs \textit{alignparts\_lmbfgs} and \textit{alignframes\_lmbfgs} are available from \\
https://sites.google.com/site/rubinsteingroup/direct-detector-align\_lmbfgs.

\section{Acknowledgements}
We thank Jianhua Zhao for providing the cryo-EM image of yeast V-ATPase used for producing figures 1 to 4 and Michael Latham and Samir Benlekbir for providing the 20S proteasome images used for the characterization of the algorithm shown in figure 5. We thank Tim Grant and Niko Grigorieff for providing data on radiation-induced fading of Fourier components ahead of publishing their manuscript and Hui Guo for characterizing an earlier version of the program. We are grateful to Alexis Rohou, Tim Grant, Niko Grigorieff, Richard Henderson, Sjors Scheres, Peter Rosenthal, and members of the Rubinstein laboratory for helpful comments on the manuscript. This work was funded by Natural Sciences and Engineering Research Council discovery grant 401724-12 (JLR) and the Canada Research Chairs program (JLR). A preprint of this manuscript was first published on arXiv.org on 24 Sept 2014 (http://arxiv.org/abs/1409.6789).

\section{Author contributions}
JLR conceived of the Fourier space objective function for optimizing particle movement estimates, wrote the programs, characterized their performance, and wrote the manuscript. MAB suggested the use of the gradient-based lm-bfgs algorithm, proposed the second order smoothing, and made other critical suggestions for the programs and the manuscript.

\bibliographystyle{elsarticle-num}
\bibliography{./Bibliography}

\begin{thebibliography}{10}
\expandafter\ifx\csname url\endcsname\relax
  \def\url#1{\texttt{#1}}\fi
\expandafter\ifx\csname urlprefix\endcsname\relax\def\urlprefix{URL }\fi
\expandafter\ifx\csname href\endcsname\relax
  \def\href#1#2{#2} \def\path#1{#1}\fi

\bibitem{mcmullan2008electron}
G.~McMullan, A.~Faruqi, Electron microscope imaging of single particles using
  the medipix2 detector, Nuclear Instruments and Methods in Physics Research
  Section A: Accelerators, Spectrometers, Detectors and Associated Equipment
  591~(1) (2008) 129--133.

\bibitem{glaeser2011images}
R.~Glaeser, G.~McMullan, A.~Faruqi, R.~Henderson, Images of paraffin monolayer
  crystals with perfect contrast: minimization of beam-induced specimen motion,
  Ultramicroscopy 111~(2) (2011) 90--100.

\bibitem{Brilot:2012zr}
A.~F. Brilot, J.~Z. Chen, A.~Cheng, J.~Pan, S.~C. Harrison, C.~S. Potter,
  B.~Carragher, R.~Henderson, N.~Grigorieff, Beam-induced motion of vitrified
  specimen on holey carbon film, J Struct Biol 177~(3) (2012) 630--7.
\newblock \href {http://dx.doi.org/10.1016/j.jsb.2012.02.003}
  {\path{doi:10.1016/j.jsb.2012.02.003}}.

\bibitem{Li:2013kq}
X.~Li, P.~Mooney, S.~Zheng, C.~R. Booth, M.~B. Braunfeld, S.~Gubbens, D.~A.
  Agard, Y.~Cheng, Electron counting and beam-induced motion correction enable
  near-atomic-resolution single-particle cryo-em, Nat Methods 10~(6) (2013)
  584--90.
\newblock \href {http://dx.doi.org/10.1038/nmeth.2472}
  {\path{doi:10.1038/nmeth.2472}}.

\bibitem{Cao:2013lq}
E.~Cao, M.~Liao, Y.~Cheng, D.~Julius, Trpv1 structures in distinct
  conformations reveal activation mechanisms, Nature 504~(7478) (2013) 113--8.
\newblock \href {http://dx.doi.org/10.1038/nature12823}
  {\path{doi:10.1038/nature12823}}.

\bibitem{Liao:2013db}
M.~Liao, E.~Cao, D.~Julius, Y.~Cheng, Structure of the trpv1 ion channel
  determined by electron cryo-microscopy, Nature 504~(7478) (2013) 107--12.
\newblock \href {http://dx.doi.org/10.1038/nature12822}
  {\path{doi:10.1038/nature12822}}.

\bibitem{Henderson:2011mz}
R.~Henderson, S.~Chen, J.~Z. Chen, N.~Grigorieff, L.~A. Passmore,
  L.~Ciccarelli, J.~L. Rubinstein, R.~A. Crowther, P.~L. Stewart, P.~B.
  Rosenthal, Tilt-pair analysis of images from a range of different specimens
  in single-particle electron cryomicroscopy, J Mol Biol 413~(5) (2011)
  1028--46.
\newblock \href {http://dx.doi.org/10.1016/j.jmb.2011.09.008}
  {\path{doi:10.1016/j.jmb.2011.09.008}}.

\bibitem{Scheres:2012yu}
S.~H.~W. Scheres, Relion: implementation of a bayesian approach to cryo-em
  structure determination, J Struct Biol 180~(3) (2012) 519--30.
\newblock \href {http://dx.doi.org/10.1016/j.jsb.2012.09.006}
  {\path{doi:10.1016/j.jsb.2012.09.006}}.

\bibitem{Scheres:2014ys}
S.~H. Scheres, Beam-induced motion correction for sub-megadalton cryo-em
  particles, Elife 3 (2014) e03665.

\bibitem{campbell20152}
M.~G. Campbell, D.~Veesler, A.~Cheng, C.~S. Potter, B.~Carragher, 2.8 {\aa}
  resolution reconstruction of the thermoplasma acidophilum 20s proteasome
  using cryo-electron microscopy, eLife 4 (2015) e06380.

\bibitem{byrd1995limited}
R.~H. Byrd, P.~Lu, J.~Nocedal, C.~Zhu, A limited memory algorithm for bound
  constrained optimization, SIAM Journal on Scientific Computing 16~(5) (1995)
  1190--1208.

\bibitem{unwin1975molecular}
P.~N.~T. Unwin, R.~Henderson, Molecular structure determination by electron
  microscopy of unstained crystalline specimens, Journal of molecular biology
  94~(3) (1975) 425--440.

\bibitem{hayward1979radiation}
S.~B. Hayward, R.~M. Glaeser, Radiation damage of purple membrane at low
  temperature, Ultramicroscopy 4~(2) (1979) 201--210.

\bibitem{Baker:2010ty}
L.~A. Baker, E.~A. Smith, S.~A. Bueler, J.~L. Rubinstein, The resolution
  dependence of optimal exposures in liquid nitrogen temperature electron
  cryomicroscopy of catalase crystals, J Struct Biol 169~(3) (2010) 431--7.
\newblock \href {http://dx.doi.org/10.1016/j.jsb.2009.11.014}
  {\path{doi:10.1016/j.jsb.2009.11.014}}.

\bibitem{grant2015measuring}
T.~Grant, N.~Grigorieff, Measuring the optimal exposure for single particle
  cryo-em using a 2.6 {\aa} reconstruction of rotavirus vp6, eLife (2015)
  e06980.

\end{thebibliography}

\end{document}